\documentclass[aps,prd,twocolumn,preprintnumbers,floatfix,nofootinbib,10pt]{revtex4-1}
\pdfoutput=1
\usepackage[english]{babel}
\usepackage{amsmath, amssymb}
\usepackage{lmodern}
\usepackage{microtype}
\usepackage[normalem]{ulem}
\usepackage{array}
\usepackage{geometry}
\usepackage{color}
\usepackage{appendix}
\usepackage{mathtools}
\usepackage{tikz}
\usepackage{multirow}
\usepackage[colorlinks=true,
    linkcolor=blue,
    filecolor=magenta,      
    urlcolor=blue,
    citecolor=blue,
    pdftitle={Sharelatex Example},
    pdfpagemode=FullScreen]{hyperref}
\usepackage{float}
\usepackage{comment}
\usetikzlibrary{positioning}
\usetikzlibrary{decorations.pathmorphing}
\usepackage{colortbl} 
\usepackage{xcolor}
\usepackage{bm} 
\definecolor{Green}{rgb}{0,0.5,0}
\usepackage{slashed}
\usepackage{hyperref}
\usepackage{physics}
\usepackage{graphicx}
\usepackage{booktabs}
\usepackage[labelfont=bf,font=small]{caption}
\usepackage{subcaption}
\usepackage{float}
\usepackage{placeins}
\usepackage{bbm}

\usepackage{siunitx}
\newcommand{\Br}{\operatorname{Br}}
\newcommand{\sigv}{\langle \sigma_{11\to22}v_{\rm rel} \rangle}

\newcommand{\keV}{\,\mathrm{keV}}

\begin{document}

\title{Sub-MeV gamma-ray lines from an endothermic interpretation of the LZ recoil candidate}

\author{Ma\'ira Dutra$^{1,2}$}
\email{mdutrava@mail.nasa.gov}

\author{Jacinto P. Neto$^{3}$}
\email{jacinto.neto.100@ufrn.edu.br}

\author{Clarissa Siqueira$^{4}$}
\email{csiqueira@on.br}

\affiliation{$^1$Southeastern Universities Research Association (SURA), 1201 New York Avenue NW, Suite 430, Washington, DC 20005, USA}
\affiliation{$^2$NASA Goddard Space Flight Center, 8800 Greenbelt Road, Greenbelt, MD 20771, USA}
\affiliation{$^3$International Institute of Physics, Federal University of Rio Grande do Norte, Campus Universit\'ario, Lagoa Nova, Natal-RN 59078-970, Brazil} 
\affiliation{$^4$Observat\'orio Nacional, Rio de Janeiro - RJ, 20921-400, Brazil}

\begin{abstract}
    Endothermic upscattering of heavy dark matter offers a compelling interpretation of the high-energy nuclear-recoil candidate recently reported by LUX-ZEPLIN, with a characteristic recoil energy $E_R \propto \delta$, where $\delta$ is the mass splitting between the dark matter states. Cosmological and experimental constraints motivate the depletion of the primordial heavier state, which can be regenerated in the Galactic halo. We show that, if the heavier state decays radiatively, the same mass splitting sets the energy of a gamma-ray line, $E_\gamma \simeq \delta$. For our benchmark, this leads to a sub-MeV line within the projected reach of the Compton Spectrometer and Imager (COSI), scheduled for launch in 2027. The confirmation of the LUX-ZEPLIN candidate event as a dark matter signal would therefore motivate dedicated searches for sub-MeV gamma-ray lines, with COSI providing a timely opportunity to test this interpretation. 
\end{abstract}

\maketitle

\section{Introduction}
\label{sec:intro}

The LUX-ZEPLIN (LZ) Collaboration has recently reported a high-energy nuclear-recoil candidate with $E_R=248 \pm 23 \,({\rm stat})\pm23\,({\rm sys})$~keV, corresponding to a momentum transfer $q=(2m_{\rm Xe}E_R)^{1/2}\simeq0.25$~GeV~\cite{LZ:2026axp}.
Although a single candidate does not establish a dark matter (DM) signal, its unusually large recoil energy motivates exploring interactions and kinematics that can populate this region of the xenon recoil spectrum. A successful interpretation must account for the candidate while remaining consistent with the absence of an accompanying excess at lower recoil energies, where conventional spin-independent elastic scattering of halo DM typically predicts a larger event rate.

Proposed interpretations include exothermic scattering~\cite{Baer:2026fpy, Xing:2026Galactic, deLima:2026shq}, fermionic DM absorption~\cite{Lou:2026idn}, boosted DM~\cite{Alhazmi:2026efz}, and cosmic-ray-boosted DM, including scenarios with momentum-dependent interactions~\cite{Heikinheimo:2026kwp, Chauhan:2026udz}. Elastic explanations have also been explored, including pseudoscalar-mediated scattering through an axion portal~\cite{Unwin:2026rdp}, a two-Higgs-doublet model supplemented by a pseudoscalar mediator~\cite{Arcadi:2026kev}, and spin-dependent $Z$ exchange in singlet-doublet Majorana DM~\cite{Elahi:2026vlm}. Neutrino-based alternatives instead attribute the recoil to atmospheric-neutrino-induced production or upscattering of new neutral fermions~\cite{Jeesun:2026vzo, Brdar:2026ukx}. Other interpretations can be found at Refs.~\cite{Chattopadhyay:2026ryw, Roy:2026wzb}.

Endothermic inelastic scattering provides a particularly direct connection between the recoil spectrum and the internal mass structure of the dark sector~\cite{Tucker-Smith:2001myb,DallaValleGarcia:2024zva}. In this framework, a lighter state $\chi_1$ scatters off a nucleus into a heavier state $\chi_2$, spending part of its initial kinetic energy to overcome the positive mass splitting $\delta \equiv m_2 - m_1>0$. The resulting kinematic threshold suppresses low-energy recoils and, for sufficiently large splittings, selects the high-speed tail of the incident DM population~\cite{Dent:2026bji, Rodd:2026Sideband, Palmisano:2026kuj,McCabe:2026crm, OHare:2026HighVelocity}. Consequently, the accepted recoil spectrum can become concentrated in the high-energy search window, motivating endothermic interpretations of the LZ candidate~\cite{Su:2026rwz, DiMauro:2026ldr,Smirnov:2026aqk, Fan:2026kxx, Freese:2026sga, Wu:2026nhi, Yin:2026jnn, Visinelli:2026kgt, Bisal:2026khf, Unwin:2026qep,Bandyopadhyay:2026gjw, Borah:2026zwf, Lee:2026jxl,Lian:2026hpm,Zhu:2026dag, Das:2026buc,Yamashita:2026ump,De:2026win,An:2026pkc,Borah:2026ris,Das:2026uyy,Chanda:2026wdl,Asadi:2026MIDB,He:2026MIDM,Cabo-Almeida:2026uqw, Barman:2026omh}. Differences in spectral shape can help discriminate inelastic scattering from boosted-DM explanations~\cite{Mahapatra:2026glu}.

\begin{figure*}[ht!]
\centering
\includegraphics[width=0.92\textwidth]{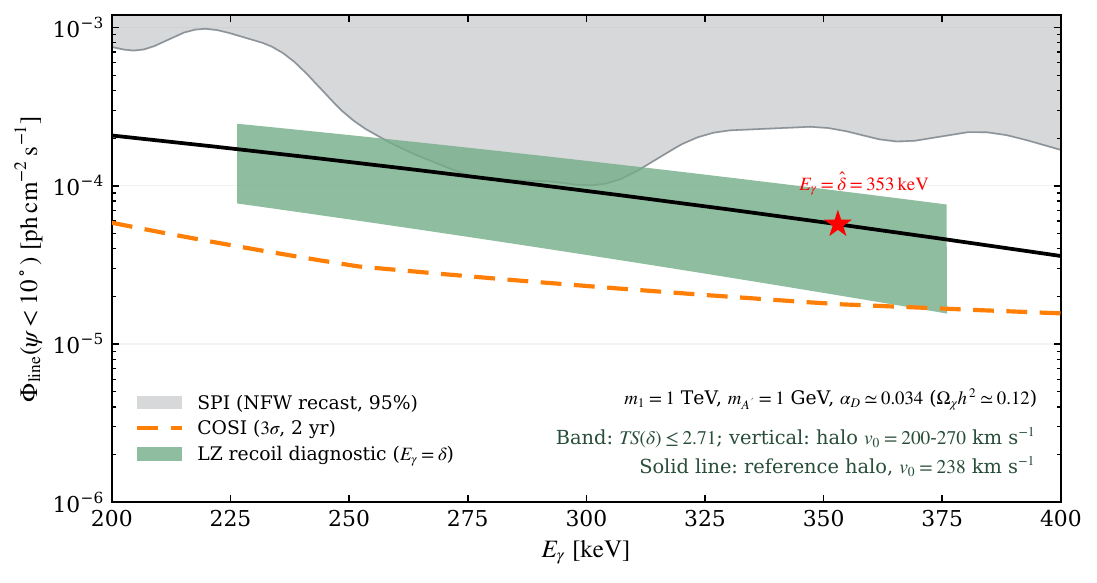}
\caption{Integrated photon line flux within $10^\circ$ of the GC for $m_1=1$ TeV, $m_{A'}=1$ GeV and $\alpha_D=0.034$. 
The green band shows splittings selected by the LZ energy diagnostic ($TS\leq2.71$), with the red star indicating the best-fit. Its vertical extent varies the emission velocity scale from $200$ to $270$ km s$^{-1}$. The solid black line is the flux for the reference halo, $v_0=238$ km s$^{-1}$. The gray region is the NFW-only $95\%$C.L. limit from Ref.~\cite{Siegert:2024hmr}. The dashed curve is the approximate $3\sigma$, two-year COSI projection~\cite{Dutra:2025COSI}.
}
\label{fig:main}
\end{figure*}

The same mass splitting that produces a high-energy nuclear recoil at the LZ experiment can also generate a distinctive indirect-detection signature.
If the excited state undergoes the radiative transition $\chi_2\to\chi_1\gamma$, the photon energy in its rest frame is
\begin{equation}
    E_\gamma=\frac{m_2^2-m_1^2}{2m_2}\simeq\delta,
\end{equation}
where the approximation holds for $\delta\ll m_1$.
Consequently, DM candidates with mass in the GeV-TeV scale can produce a sub-MeV gamma-ray line whose energy measures the splitting between the dark states. This scenario was motivated by the possibility of explaining the $3.5$~keV X-ray line along with the GeV Galactic Center excess and DM signatures at colliders~\cite{Falkowski:2014sma, Lee:2015sha, Borah:2015rla}. Here, \textit{we propose a connection between nuclear-recoil kinematics and gamma-ray spectroscopy}.

For splittings of a few hundred keV, the associated line falls within the $0.2$--$5$~MeV energy range of the Compton Spectrometer and Imager (COSI)~\cite{Tomsick:2023aue}. COSI is scheduled for launch in 2027 and will be an excellent tool to search for gamma-ray lines, since its energy resolution is of about 1\% at the MeV scale. Moreover, COSI will improve the current limits in the MeV range due to its large instantaneous field-of-view ($>25\%$ of the sky) and better background discrimination. For a recent forecast of COSI limits on gamma-ray lines due to dark matter decay, see Ref.\,\cite{Anthony-Petersen:2026oxk}.

In this work, we investigate a mechanism in which the excited-state cosmological population is depleted, but Galactic regeneration repopulates it via DM self-interactions, $\chi_1 \chi_1 \to \chi_2 \chi_2$~\cite{Finkbeiner:2014sja, Schutz:2014nka, Berlin:2023qco}. The subsequent radiative decay of $\chi_2$ yields a gamma-ray line. For relevant regions of the parameter space, the regeneration and decay rates reach local equilibrium, making the gamma-ray line flux determined by the excitation rate and the decay branching ratio. The observable flux then depends on the square of the DM density and on the velocity distribution, with particular sensitivity to the high-speed particles that can overcome the excitation threshold. We focus on gamma-ray lines coming from the Galactic center (GC).

We assess this mechanism in a pseudo-Dirac dark sector coupled to a kinetically mixed dark photon~\cite{Zhu:2026dag}. Dark photon exchange mediates both endothermic nuclear scattering and collisional excitation in the halo, while annihilation into dark photons governs thermal freeze-out.
We supplement these interactions with an electromagnetic transition dipole that permits $\chi_2\to\chi_1\gamma$, taking its contribution to freeze-out and nuclear scattering to be subdominant.
The dipole thus depletes the primordial excited population and converts halo excitations into line photons.

Our main results are summarized in Fig.~\ref{fig:main}.
We calculate the line flux within a disk of $10^\circ$ around the GC (solid black line), treating halo excitation through a coupled-channel scattering calculation, and compare it with existing INTEGRAL/SPI limits~\cite{Siegert:2024hmr} (gray region) and the projected COSI sensitivity~\cite{Dutra:2025COSI} (dashed orange curve).
For a benchmark with $m_1=1$~TeV, $m_{A'}=1$~GeV, and $\alpha_D=0.034$, which reproduces the correct relic abundance, we identify mass splittings selected by an approximate diagnostic of the LZ recoil energy (green band) and examine the dependence of the predicted emission on the halo velocity distribution (vertical extent). The red star in Fig.~\ref{fig:main} represents the best-fit for the mass splitting, $\delta=353$~keV. Notably, this benchmark yields a line within the projected COSI reach under the adopted halo assumptions. These results demonstrate that sub-MeV line searches can test an endothermic interpretation of the LZ candidate featuring a radiatively decaying excited state.

\begin{figure*}[ht!]
\centering
\includegraphics[width=0.92\textwidth]{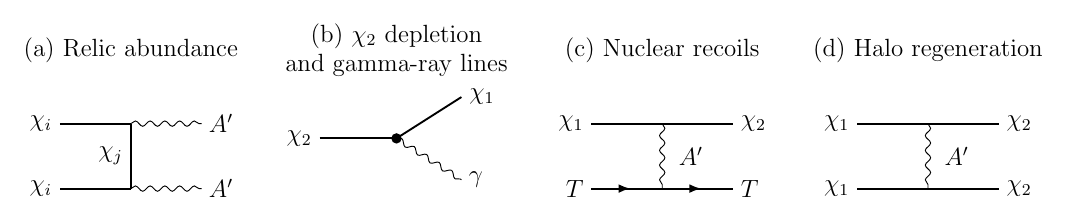}
\caption{Feynman diagrams illustrating the main physical processes at play: (a) dark matter annihilation into dark photons sets the relic abundance; (b) a dipole operator depletes the primordial $\chi_2$ population, enabling gamma-ray lines at energy $E_\gamma \sim \delta$; (c) endothermic nuclear recoils at energies $E_R^\star \propto \delta$ occur via dark photon exchange; and (d) excited state regeneration in the Galactic halo.}
\label{fig:diagrams}
\end{figure*}

The remainder of this paper is organized as follows.
Section~\ref{sec:model} introduces the dark-sector framework and its nuclear-recoil phenomenology. Section~\ref{sec:LZComp} analyses the compatibility of our mechanism with the LZ event candidate using an approximate energy-only likelihood.  
Section~\ref{sec:ID} describes the regeneration of excited states in the Galactic halo and the resulting gamma-ray line.
We present our results and discussion in Sec.~\ref{sec:results}, before concluding in Sec.~\ref{sec:conclusions}.

\section{Endothermic interpretation}
\label{sec:model}

We consider two Majorana states forming a pseudo-Dirac pair, $\chi_1$ and $\chi_2$, with masses $m_1$ and $m_2=m_1+\delta$. Their interaction with a massive dark photon $A'$ is off-diagonal, allowing $\chi_1$ to scatter into $\chi_2$. Kinetic mixing with the photon connects this dark sector to electrically charged Standard Model particles~\cite{Zhu:2026dag}.
The relevant low-energy interactions are
\begin{equation}
    \mathcal L_{\rm int}\supset  i g_D A'_\mu\bar\chi_2\gamma^\mu\chi_1 + \epsilon e A'_\mu J_{\rm EM}^\mu,
\label{eq:vector}
\end{equation}
where $g_D$ is the coupling strength of the new gauge interaction carried by $A'$ and $\epsilon$ is the kinetic mixing. We also define a useful coupling strength quantity $\alpha_D=g_D^2/(4\pi)$. In the regime $m_{A'} \ll m_Z$, we can safely neglect the contribution of a $Z$-boson exchange. 

Both $\chi_i$ states participate in the freeze-out production when the mass splitting is much smaller than the freeze-out temperature, $\delta\ll T_f$. In the secluded regime $m_{A'}<m_1$, the leading channels include $\chi_i\chi_i\to A'A'$, illustrated in the panel (a) of Fig.~\ref{fig:diagrams}. Schematically, their cross sections scale as $g_D^4/m_1^2$. The abundance must therefore be calculated considering both states and all relevant channels, including coannihilations. We use \texttt{micrOMEGAs} to perform these numerical calculations~\cite{Belanger:2026asz}. 

For $\delta<2m_e$, the $\chi_2\to\chi_1 e^-e^+$ decay channel is kinematically forbidden, leaving $\chi_2\to\chi_1+3\gamma$ and $\chi_2\to\chi_1\bar{\nu}\nu$ as the leading decay channels in the minimal dark photon model.
Since the former is quite slow, such a minimal dark photon framework alone can leave a long-lived excited population, leading to strong constraints from the cosmic microwave background and enabling exothermic scatterings~\cite{Slatyer:2015jla, Slatyer:2016qyl, Berlin:2023qco, Zhu:2026dag}.
A well-motivated way to deplete that primordial population is to consider an electromagnetic transition moment~\cite{Feldstein:2010su, Chang:2010en, Weiner:2012gm, Giunti:2014ixa, Eby:2023wem}
\begin{equation}
    \mathcal L_{\rm dip}=\frac{i \mu_{12}}{2} \bar\chi_ 2\sigma^{\mu\nu}\chi_1 F_{\mu\nu} ,
\label{eq:dipole}
\end{equation}
where $\mu_{12}$ corresponds to the magnetic dipole moment and $F_{\mu\nu}$ is the electromagnetic field-strength tensor
\footnote{We use the explicitly Hermitian Majorana convention of Ref.~\cite{Eby:2023wem}, where the transition tensor bilinear is antisymmetric under $\chi_1\leftrightarrow\chi_2$, so the factor $i$ makes the last term Hermitian without adding another identical Hermitian-conjugate term. In this convention $\mu_{12}$ is real and signed, with mass dimension $-1$.}.

This operator leads to the radiative decay, $\chi_2 \to \chi_1 \, \gamma$, depleting the excited state $\chi_2$ after freeze-out. Although the dipole moment operator affects the population of the excited states, its contribution to the freeze-out production and to the DM-nucleon scattering is negligible (See Appendix~\ref{app:direct_detection}). Hence, the expected LZ signal comes solely from the minimal dark photon scenario. 

For a target isotope of mass $m_T$, to leading order in $\delta/m_1$, the minimum incoming speed  required to produce a recoil energy $E_R$ is~\cite{Tucker-Smith:2001myb, Barello:2014}
\begin{equation}
    v_{\min}(E_R)=\frac{1}{\sqrt{2 m_T E_R}} \left( \frac{m_T E_R}{\mu_{\chi T}} + \delta \right),
\label{eq:vmin}
\end{equation}
where $\mu_{\chi T} = m_1 m_T/(m_1+m_T)$ is the reduced mass. Therefore, the endothermic scattering, Fig.~\ref{fig:diagrams}(c), requires the extra energy to produce the heavier state, suppressing low-energy events and selecting the fast part of the halo distribution. Additionally, requiring $v_{\rm min} (E_R) \leq v_{\rm max}$, we obtain a bound for the maximum mass splitting,
\begin{equation}
    \delta_{\rm max}(m_1) = \sqrt{ 2m_T E_R}\left(v_{\rm max} - \sqrt{\dfrac{m_T E_R}{2 \mu_{\chi T}^2}}  \right),
\end{equation}
which gives
\begin{equation}
    \delta_{\rm max} = 406.6\;{\rm keV}\;\left(1 - \frac{74.4\;{\rm GeV}}{m_1} \right)
\end{equation}
for $^{131}$Xe at the candidate recoil energy $E_R = 248$ keV, assuming a maximum detector-frame speed $v_{\max}\simeq 798$ km s$^{-1}$~\cite{OHare:2026HighVelocity}. Under this assumption, no positive splitting is kinematically compatible with this recoil for $m_1 \leq 74.4$ GeV, while $\delta_{\max}$ approaches $406.6$ keV in the heavy-DM limit.

\begin{figure}[t!]
    \centering
    \includegraphics[width=\linewidth]{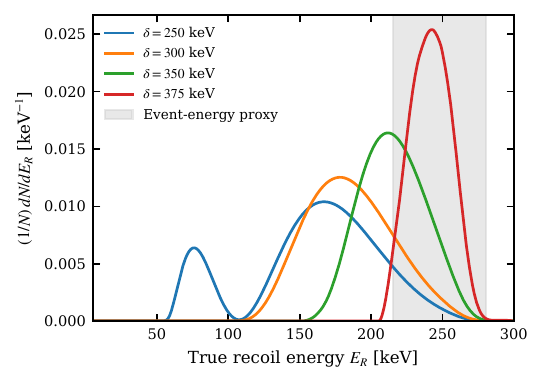}
    \caption{Xenon recoil spectra, each normalized to unit area, for the masses in Fig.~\ref{fig:main} and different mass splittings. The shaded gray region indicates the event-energy proxy $248 \pm \sqrt{23^2+23^2}$ keV.}
\label{fig:DD_norm_spectrum}
\end{figure}

Fig.~\ref{fig:DD_norm_spectrum} shows normalized spectra as a function of the recoil energy $E_R$ for different mass splittings. Notably, changing the mass splitting modifies the spectral shape, both the band and the peak. That happens because the recoil energy is confined to a band at a fixed incoming speed, 
\begin{align}
    E_R^{\pm}(v) = \frac{\mu_{\chi T}^2}{2 m_T} \left( v \pm \sqrt{v^2  - v_\delta^2} \right)^2, \quad v_\delta = \sqrt{\frac{2 \delta }{\mu_{\chi T}} } \,,
\end{align}
where $v_\delta$ is the threshold speed, and the spectrum is empty for $v < v_\delta$ and collapses to a single point at $v = v_\delta$,
\begin{equation}
    E_R^\star = \frac{\mu_{\chi T}\delta}{m_T}\, ,
\end{equation}
which corresponds to the recoil energy at which $v_{\min}$ is minimized. Note that it varies for different isotopes.

We calculate the xenon response with WimPyDD~\cite{Jeong:2021bpl}, including the finite mediator propagator and LZ selection efficiency. The leading interaction is coherent, with strength proportional to $\epsilon g_D$. 

\section{Compatibility with the LZ candidate}
\label{sec:LZComp}

We estimate the splitting range compatible with the LZ candidate using an approximate energy-only likelihood, following the phenomenological approaches of Refs.~\cite{Zhu:2026dag, Dent:2026bji}. Assuming that the candidate recoil is a signal and neglecting backgrounds, the observed counts $n_i$ given the expected counts $\lambda_i$ in an energy bin $i$ are given by
\begin{equation}
    n_i \sim {\rm Poisson}(\lambda_i) = \frac{\lambda_i^{n_i} e^{-\lambda_i}}{n_i!}\,,
\end{equation}
with $\lambda_i (\epsilon,\delta) = A(\epsilon)\,s_i^{\rm ref}(\delta)$ and $s_i^{\rm ref}$ the reference signal counts. The kinetic mixing determines the signal normalization $A(\epsilon)=(\epsilon/\epsilon_{\rm ref})^2$, where we take $\epsilon_{\rm ref}=10^{-5}$, whereas the mass splitting determines the spectral shape.

With one event observed in the entire main window of nuclear recoils ($E_R = 5.4 - 300$~keV) and no event observed in the sideband ($E_R = 350-680$~keV), the logarithm of the likelihood is given by
\begin{align}
    \log \mathcal{L}(\epsilon,\delta) &= \sum_i (n_i \log \lambda_i-\lambda_i-\log(n_i!)) \nonumber \\
    &= \sum_{j \,\in\, \rm main}\log \lambda_j - \sum_i\lambda_i \nonumber \\
    &= \log(AW) - A(U+V)\,.
\end{align}

In the equation above, $U(\delta)$ is the number of accepted counts in the main window,
\begin{equation}
    U(\delta) = \int_{5.4\rm keV}^{300\rm keV}dE_R \frac{dN}{dE_R}\Big|_\delta\,,
\end{equation}
$V(\delta)$ is the number of accepted counts in the sideband, assuming a constant efficiency of $96\%$~\cite{Dent:2026bji},
\begin{equation}
    V(\delta) = \int_{350\rm keV}^{680\rm keV}dE_R \frac{dN}{dE_R}\Big|_\delta\,,
\end{equation}
and $W(\delta)$ is the number of accepted counts in the main window convolved with an approximate detector response,
\begin{equation}
    W(\delta) = \frac{dN_{\rm obs}}{dE_{\rm obs}}\Big|_\delta = \int_{5.4\rm keV}^{300\rm keV}dE_R \frac{dN}{dE_R}\Big|_\delta G(E_{\rm obs|E_R})\,.
\end{equation}

Since the observed recoil energy was $E_{\rm obs} = 248 \pm 23(\rm stat)\pm23(\rm sys)$ keV, we consider the approximate response as a Gaussian with uncertainty $\sigma_E=\sqrt{23^2+23^2}$~keV.

\begin{figure*}[ht!]
    \centering
    \includegraphics[width=0.98\textwidth]{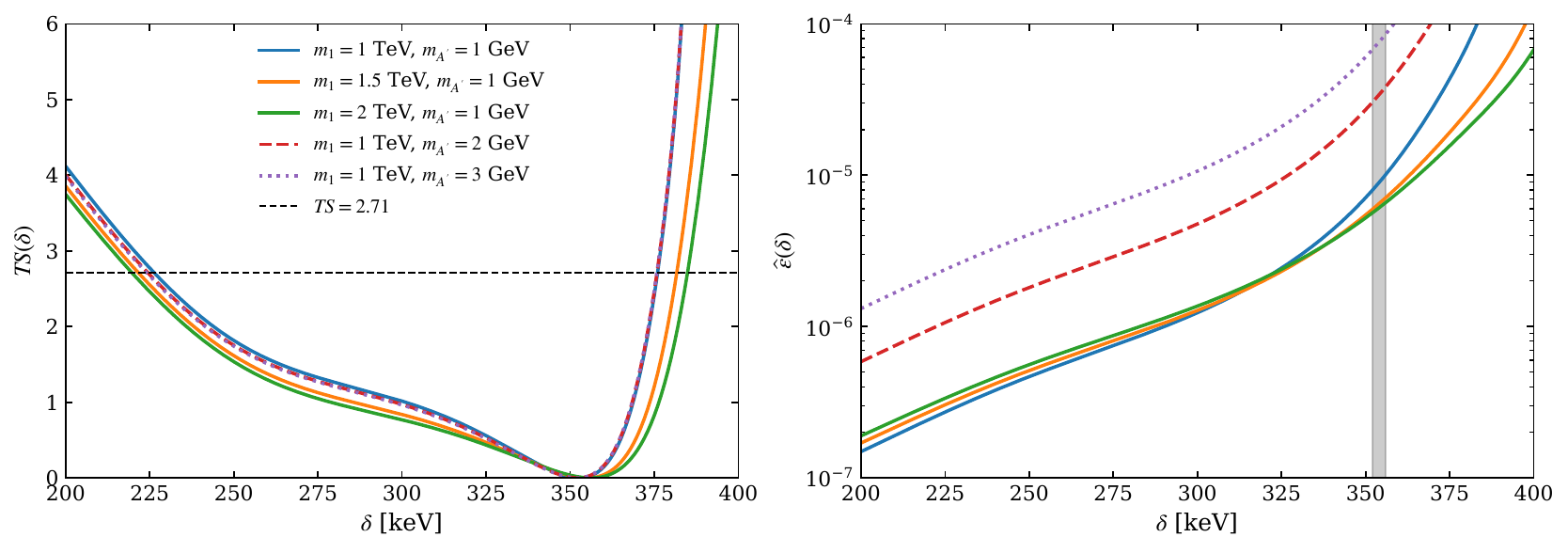}
    \caption{Illustrative criterion for agreement with the LZ candidate considered in this work, with the mass splitting $\delta$ determining the spectral shape and the kinetic mixing parameter $\epsilon$ determining the signal normalization. Our approximate likelihood model has one event in the main window, no event in the sideband region, and assumes zero background counts. We show the two-sided test statistic (left) and the profiled $\epsilon$ (right) as a function of $\delta$ for different values of DM and dark photon masses. The best-fit values for $\delta$ are found in the range $352-356$~keV and are indicated in the gray vertical band in the right panel.}
\label{fig:LZdiagnostic}
\end{figure*}

We profile over the kinetic mixing and use as an illustrative criterion of compatibility with the LZ candidate a test statistic below the asymptotic threshold for a two-sided interval at $90\%$~C.L.~\cite{Cowan:2010js}. The test statistic is a function of the mass splitting and is given by
\begin{equation}
    \rm TS (\delta) = -2\log\frac{\mathcal{L}(\delta,\hat\epsilon(\delta))}{\mathcal{L}(\hat\delta,\hat\epsilon(\hat\delta))}\,,
\end{equation}
with hat indicating best-fit values. In this simple analytical test statistic, $\hat A (\delta) = 1/(U(\delta)+V(\delta))$, so $\hat \epsilon(\delta) = \epsilon_{\rm ref} \sqrt{\hat A(\delta)}$. Thus, values of $\delta$ for which $\rm TS(\delta)<2.71$ are regarded as compatible with the LZ candidate. This determines the horizontal extent of the green band in Fig.~\ref{fig:main}. However, we emphasize that this is not a statistical confidence region.

In Fig.~\ref{fig:LZdiagnostic}, we show how the test statistic (left) and the best-fit of the kinetic mixing (right) depend on the mass splitting for different DM and dark photon masses. The best-fit values for $\delta$ are found in the range $352-356$~keV, with $\hat \delta = 353$~keV for our benchmark $m_1=1$~TeV and $m_{A'}=1$~GeV. Heavier mediators require larger $\epsilon$ to compensate the suppression in the scattering signal. 

\section{Gamma-ray line signal}
\label{sec:ID}

As we discussed in Section \ref{sec:model}, the dark matter relic is composed of $\chi_1$ states while the primordial $\chi_2$ population has decayed via $\chi_2\to\chi_1\gamma$ after freeze-out. In this scenario, the secluded annihilation of $\chi_1$ can lead to indirect detection signal like continuum gamma-rays ($4f$)~\cite{Pospelov:2007mp}, and the process $\chi_1\chi_1\to\gamma\gamma$, enabled by the dipole operator, can lead to TeV-scale gamma-ray lines~\cite{Wu:2026nhi}. However, these signals are not expected to be strong in our case~\cite{Elor:2015tva,Profumo:2017obk}.

Interestingly, the decay of the primordial population of $\chi_2$ does not eliminate an astrophysical signal. Although the lifetime is too short to preserve excited particles from the early universe, the same vector interaction can produce them today through $\chi_1\chi_1 \to \chi_2 \chi_2$, see Fig.~\ref{fig:diagrams}(d)~\cite{Finkbeiner:2014sja, Berlin:2023qco, Xing:2026Galactic}, so their subsequent radiative decay (Fig.~\ref{fig:diagrams}(b)) can generate unsuppressed gamma-ray lines. 

By energy conservation, we must have $k_1^2/m_1 > 2\delta$ to produce two excited states in the halo, with $k_1=m_1v_{\rm rel}/2$ the momentum of an incident particle $\chi_1$. Thus, the minimum relative velocity allowing $\chi_1\chi_1\to\chi_2\chi_2$ is
\begin{align}
    \label{eq:speed-exc-threshold}
    v_{{\rm rel}} > v_{\rm th} &=\sqrt{\frac{8\delta}{m_1}}\nonumber \\ 
    & \simeq 502\ {\rm km\,s^{-1}} \left(\frac{\delta}{350\keV}\frac{1\ {\rm TeV}}{m_1}\right)^{1/2}.
\end{align}

The excitation rate, $\Gamma_{\rm exc} = n_1^2 \sigv$, with $n_1$ the number density of $\chi_1$, is consequently sensitive to the halo velocity distribution. We calculate $\sigma_{11\to 22}$ by solving the coupled-channel scattering problem, since a perturbative treatment is not sufficient for our benchmarks, $\alpha_D \,m_1/m_{A'} > 1$~\cite{Slatyer:2009vg, Schutz:2014nka} (see Appendix \ref{sec:excited}). Considering each incoming velocity drawn from a Maxwellian distribution truncated at the scape velocity $v_{\rm esc}$, the average over relative velocities is given by
\begin{align}
\sigv &= \int_{v_{\rm th}}^{2v_{\rm esc}} dv_{\rm rel} f_{\rm rel}(v_{\rm rel}) \sigma_{11\to 22} (v_{\rm rel}) v_{\rm rel} \nonumber \\
& = \int_{v_{\rm th}}^{2v_{\rm esc}} dv_{\rm rel} \frac{4v_{\rm rel}^2}{\pi v_0^4 N^2} \sigma_{11\to 22} (v_{\rm rel}) v_{\rm rel} \nonumber \\
& ~~ \times \left[ \frac{\sqrt{\pi}}{2\sqrt{2}} v_0 {\rm erf}\left(\frac{\sqrt{2}h}{v_0}\right) e^{-v_{\rm rel}^2/(2v_0^2)} \right. \nonumber \\
& ~~~~~~~ \left. - \frac{v_0^2}{2v_{\rm rel}} \left(e^{2h v_{\rm rel}/v_0^2}-1\right) e^{-2v_{\rm esc}^2/v_0^2} \right]\,,
\label{eq:avsigmav}
\end{align}
where $h=v_{\rm esc}-v_{\rm rel}/2$ quantifies the truncation ($h=0$ for $v_{\rm rel}=2v_{\rm esc}$) and $N={\rm erf}(z)- 2z e^{-z^2}/\sqrt{\pi}$, with $z=v_{\rm esc}/v_0$, is the normalization of the truncated Maxwellian distribution of one incident velocity, with $\rm erf$ the error function. Our reference halo has $v_0=238$ km s$^{-1}$ and $v_{\rm esc}=544$ km s$^{-1}$.

After production, these excited state particles promptly decay into monochromatic photons through the electromagnetic dipole operator in Eq.~\eqref{eq:dipole},
\begin{equation}
    \Gamma_{\chi_2\to\chi_1\gamma} \simeq \frac{|\mu_{12}|^2}{\pi}E_\gamma^3 \,,
\label{eq:width}
\end{equation}
with the photon energy given by
\begin{align}
    E_\gamma = \frac{m_2^2-m_1^2}{2m_2} = \delta\left(\frac{m_1}{m_2}+\frac{\delta}{2m_2}\right) \simeq \delta \, .
\end{align}

The number of photons per unit of time and volume is given by $\Gamma_{\rm emis} = n_2 \Gamma_{\chi_2\to\chi_1\gamma}$, with $n_2$ the number density of $\chi_2$. When production and decay reach local equilibrium ($\dot{n_2}=\Gamma_{\rm exc}-n_2 \Gamma_2 \simeq 0$), with $\Gamma_2$ the total decay width of $\chi_2$, we have $n_2 \simeq n_1^2\sigv/\Gamma_2$. Assuming $\delta=353$ keV and $v_0=238$~km~s$^{-1}$, the fraction of excited states is given by
\begin{align}\label{eq:f2-def}
    f_2 &\equiv \frac{n_2}{n_1+n_2} \simeq \frac{n_2}{n_1} \nonumber\\
    &\simeq 4\times 10^{-16} \frac{\rho_1}{0.4 \rm GeV/cm^3} \frac{1 \rm TeV}{m_1} \frac{\langle \sigma_{11\to 22} v\rangle }{10^{-20}\rm cm^3/s} \frac{\tau_2}{10^8 \rm s}.
\end{align}

The integrated flux is therefore
\begin{equation}
    \Phi_{\rm line} = \frac{\Br_\gamma}{4\pi m_1^2} \sigv J,
\label{eq:flux-general}
\end{equation}
with $J = \int_{\Delta\Omega}d\Omega\int ds\,\rho_1^2(r)$ the J-factor, $\Br_\gamma=\Gamma_{\chi_2\to\chi_1\gamma}/\Gamma_{\rm tot}$ the branching ratio, and $\rho_1(r)$ the dark matter density of $\chi_1$.

Thus, the gamma-ray line flux is fixed by the excitation rate and can be sizable even if very few excited particles are present at any instant. 

\begin{figure*}
    \centering
    \includegraphics[width=\textwidth]{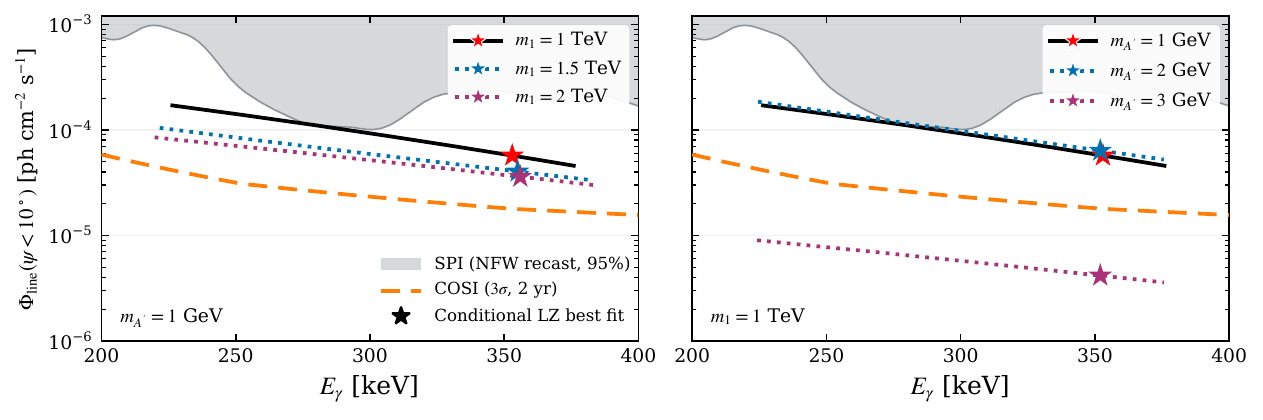}
    \caption{Dependence of the line flux on the DM mass (left) and mediator mass (right), at fixed $\alpha_D=0.034$ and reference halo $v_0=238$ km s$^{-1}$. Each curve is restricted to $TS\leq2.71$ after profiling $\epsilon$ for that mass choice; the endpoints encode the recoil selection, with stars marking the best-fit values. The SPI region and COSI projection follow Fig.~\ref{fig:main}. The common halo density is held fixed to isolate the parameter dependence.
    }
\label{fig:mass-dependence}
\end{figure*}

\section{Results and discussion}
\label{sec:results}

With the main ingredients at hand, we can now discuss our main results. In Fig.~\ref{fig:main} we show the predicted gamma-ray line flux integrated over a $10^\circ$ disk around the GC (black curve), given by Eq.~\eqref{eq:flux-general}, as a function of the line energy $E_\gamma \approx \delta$. It was calculated for the benchmark parameters $m_1  = 1\;{\rm TeV}$, $m_{A'} = 1\;{\rm GeV}$, and $\alpha_D  = 0.034$, which gives the correct relic abundance $\Omega_\chi h^2 \simeq 0.12$, and for a reference halo velocity scale $v_0 = 238$~km~s$^{-1}$. 

The green band, representing an approximate criterion for compatibility with the LZ candidate, is twofold. On the one hand, the horizontal extent shows the selected mass splittings $\delta$, as discussed in Section~\ref{sec:LZComp}. The best-fit value $\hat\delta \simeq 353$~keV is indicated as a red star. On the other hand, the vertical extent shows how the line flux varies as $v_0$ changes from $200$ to $270$ km s$^{-1}$. Changing $v_0$ modifies the relative velocity distribution $f_{\rm rel}(v_{\rm rel})$ and thus the halo velocity average shown in Eq.~\eqref{eq:avsigmav}. The local velocity distribution used to calculate nuclear recoils (see Eq.~\eqref{eq:dd_rate}) is held fixed throughout this variation.

The gray region is excluded by the current limit on gamma-ray lines from direct DM annihilation into two photons from 16 years of INTEGRAL/SPI observations~\cite{Siegert:2024hmr}. In our numerical results, we consider a Navarro-Frenk-White (NFW) profile with $\rho_\odot=0.4$ GeV cm$^{-3}$ and rescale this limit to a $10^\circ$ GC disk, which gives $J\simeq 4\times10^{22}$ GeV$^2$ cm$^{-5}$. The dashed curve is the $3\sigma$, two-year COSI projection derived in Ref.~\cite{Dutra:2025COSI}. This projection rescales the point source sensitivity reported in Ref.~\cite{Tomsick:2023aue} to a $10^\circ$ GC disk. 

As discussed in the previous section, under the assumption of local equilibrium between excitation and decay, the line flux depends only on the branching ratio ${\rm BR}_\gamma$ and not on the partial decay given in Eq.~\eqref{eq:width}. Thus, the actual value of the dipole momentum $\mu_{12}$ is not relevant in our numerical results. However, it is worth estimating the parameter space behind this assumption. For line energies in the range $200-400$ keV, lifetimes in the range $\tau_2 = 10^4 - 10^{12}$~s correspond to dipole moments of order $|\mu_{12}| \simeq \times 10^{-9} - 10^{-13} \, \mathrm{GeV}^{-1}$. As benchmark for our analysis, we take $\tau_2 = 10^8$~s, with $|\mu_{12}| \simeq 2.168 \times 10^{-11} \, \mathrm{GeV}^{-1}$. For the chosen parameters, the dipole contribution is taken to dominate over the other decay channels, such that $\rm BR_\gamma \simeq 1$. On the other hand, the dipole contribution to xenon recoils is small compared with the vector-mediated transition (see Appendix \ref{app:direct_detection}). Moreover, the short propagation distance, approximately $8 \times 10^{-4}$~pc at our reference speed, justifies treating the decay as local.

A possible consequence of radiative de-excitation is a recoil followed by a photon inside the detector~\cite{Lin:2010zz}, as discussed in magnetic inelastic interpretations of LZ~\cite{Asadi:2026MIDB, He:2026MIDM}. Related searches exploit upscattering in the Earth followed by decay inside a detector~\cite{Eby:2023wem}. For our lifetime, the decay probability over a path $L$ is $P_{\rm dec}\simeq L/(v_2\tau_\gamma)\simeq3\times10^{-14}$ for $L=1$ m and an excited-state speed $v_2 = 300$ km s$^{-1}$. Decays inside LZ therefore do not modify our isolated-recoil prediction.

The recoil and photon signals constrain different combinations of parameters. At fixed masses, nuclear scattering fixes approximately $\epsilon g_D$, whereas excitation in the halo depends separately on $\alpha_D$, $m_{A'}$, and the available velocities.  

In Fig.~\ref{fig:mass-dependence}, we show this dependence explicitly by calculating the gamma-ray flux at the accepted mass splittings for different masses of $\chi_1$ (left) and $A'$ (right). Changing $m_1$ lowers the excitation threshold but also reduces the number density entering the flux, as can be observed in the left panel. On the other hand, changing $m_{A'}$ modifies the non-perturbative scattering cross section in a non-monotonic way. For heavy enough $A'$, as illustrated for the case $m_{A'}=3$~GeV, the predicted flux might lie below the COSI projection while retaining the recoil interpretation and the correct relic abundance. Improving MeV-scale indirect detection limits can therefore further probe this interpretation.

The results presented in this work are timely and robust. Since LZ detected a candidate signal at high-energy recoil, which has no explanation so far, and COSI is scheduled for launch next year, the possibility of narrowing the first DM signal is knocking at the door. 
The scenario considered here can be applied to other framework realizations and can be explored in a complementary way using different experiments and search strategies. For instance, complementary constraints arise from DM capture and subsequent annihilation in the Sun: neutrino observations can strongly restrict both elastic and inelastic interpretations, depending on their scattering interactions, annihilation channels, and the evolution of the captured population~\cite{Pospelov:2026ewn, Bose:2026ndd, Nguyen:2026lui, Bose:2026szs, Capolupo:2026uyi}.

\section{Conclusions}
\label{sec:conclusions}

We have shown that one of the main interpretations of the high-energy recoil event candidate recently reported by LZ, the endothermic interpretation, can be accompanied by a sub-MeV gamma-ray line when the excited DM state is regenerated in the Galactic halo and decays radiatively. 

As a well-motivated realization of the endothermic interpretation, we considered pseudo-Dirac DM upscattering through the exchange of a light dark photon. The same kinetically mixed dark photon enables the regeneration of the excited state in the Galactic halo. Its subsequent decay via a dipole transition operator produces a potentially detectable gamma-ray line. The mass splitting sets both the characteristic recoil-energy scale, $E_R^\star=\mu_{\chi T}\delta/m_T$, and the gamma-ray line energy, $E_\gamma\simeq\delta$.

For our main benchmark (see Fig.~\ref{fig:main}), the resulting gamma-ray line flux lies within the projected sensitivity of the COSI telescope, scheduled for launch in 2027. COSI will probe yet unexplored regions of the DM parameter space, especially for line searches. The predicted line flux depends on the dark-sector parameters and halo velocities. This makes COSI complementary to searches for high-energy recoils in direct detection experiments, even if the LZ candidate is not confirmed, and motivates a joint analysis of the two signals.

\section*{Acknowledgements}
MD is supported through a cooperative agreement with the Center for Research and Exploration in Space Sciences and Technology II (CRESST II) between NASA Goddard Space Flight Center and University of Maryland, College Park, under award number 80GSFC24M0006.
JPN is supported by the Simons Foundation (Award Number:1023171-RC) and the Fundação de Amparo e Promoção da Ciência, Tecnologia e Inovação do Rio Grande do Norte (FAPERN) via the call number 06/2026 - FAPERN/IIP, with process number 10910022.000493/2024-04. 
CNPq supports CS through grant numbers 304944/2025-4 and 406718/2025-3. CS is supported by the São Paulo Research Foundation (FAPESP) through grant number 2021/01089-1.
We acknowledge the use of OpenAI's ChatGPT and Codex to assist with numerical implementation, consistency checks, literature searches, and to improve the wording in the manuscript. The scientific conclusions and results are of our entire responsibility.

\bibliographystyle{JHEP}
\bibliography{refs}

\appendix

\section{Direct  detection details}
\label{app:direct_detection}

We interpret the LZ recoil event as arising from the endothermic transition $\chi_1 T \to \chi_2 T$, where $T$ denotes a xenon isotope and the mass splitting $m_2 - m_1 = \delta > 0$. We assume that the local DM density is dominated by $\chi_1$, following the depletion of the primordial excited-state population $\chi_2$. 

For $m_{A'}\ll m_Z$, dark photon exchange couples the transition current to the electromagnetic charge of the nucleus (Fig.~\ref{fig:diagrams}c). At leading order in the non-relativistic expansion, the interaction is described by the spin-independent operator $\mathcal{O}_1 = \mathbf{1}_\chi \, \mathbf{1}_N$~\cite{DelNobile:2021wmp}, with nucleon coefficients
\begin{equation}\label{eq:dd_coefficients}
    c_1^p(q)=\frac{g_D\epsilon e}{m_{A'}^2+q^2}, \qquad c_1^n (q) = 0 \,,
\end{equation}
due to electric charge combinations of the up- and down-quark coefficients.
The differential cross section is therefore, within the kinematically accessible recoil interval,
\begin{equation}\label{eq:dd_cross_section}
    \frac{d\sigma_T}{dE_R} =\frac{8\pi\alpha\alpha_D\epsilon^2 Z_T^2 m_T}{v^2\bigl(m_{A'}^2+2m_T E_R\bigr)^2}F_{p,T}^2(q) \Theta(v - v_{\rm min}),
\end{equation}
with $\alpha_D = g_D^2/(4\pi)$ and $\alpha = e^2 /(4\pi)$. Here $Z_T = 54$ denotes the number of protons for xenon, and $F_{p, T}^2$ is the proton charge response normalized to unity at the zero-momentum transfer limit. Numerically, we evaluate this response using the isotope-dependent $W_M$ functions supplied by \texttt{WimPyDD}, including the natural xenon abundances~\cite{Jeong:2021bpl}. The full mediator propagator is retained throughout, which is relevant at the candidate energy, where $q\simeq0.25$~GeV, and the nuclear response can exhibit substantial diffraction structure.

The rate per unit detector mass is
\begin{equation} \label{eq:dd_rate}
 \frac{dR}{dE_R} =\frac{\rho_\chi}{m_1}\sum_T\frac{\xi_T}{m_T} \int_{v\geq v_{\min}^T}\!d^3v\, f_{\mathrm{lab}}(\mathbf v)\,v\, \frac{d\sigma_T}{dE_R}, 
\end{equation}
where $\xi_T$ is the isotope mass fraction and
$f_{\mathrm{lab}}$ is normalized to unity. For
Eq.~\eqref{eq:dd_cross_section}, the velocity dependence is encoded in the mean inverse speed
\begin{equation}\label{eq:dd_eta}
    \eta(v_{\min})=\int_{v\geq v_{\min}}d^3v\, \frac{f_{\mathrm{lab}}(\mathbf v)}{v},
\end{equation}
and therefore, we can write
\begin{equation} \label{eq:rate-vector-natural}
 \frac{d R}{d E_R} = \frac{\rho_\chi}{m_1}\sum_T \xi_T \frac{8\pi\alpha\alpha_D\epsilon^2 Z_T^2 F_{p,T}^2(q)}{\bigl(m_{A'}^2+ 2m_T E_R\bigr)^2}\, \eta(v_{\min})\,.
\end{equation}
We use a truncated Maxwellian Galactic distribution with $v_0=238\,\mathrm{km\,s^{-1}}$ and $v_{\mathrm{esc}}=544\,\mathrm{km\,s^{-1}}$, shifted by a fixed laboratory speed $v_{\mathrm{lab}}=253.7\,\mathrm{km\,s^{-1}}$. Annual modulation is neglected in this time-averaged treatment. The maximum velocity of the resulting distribution is $v_{\max}=v_{\mathrm{esc}}+v_{\mathrm{lab}}\simeq798\,\mathrm{km\,s^{-1}}$. We evaluate $\eta$ continuously near this endpoint. For consistency with the Galactic density model, we take $\rho_\chi=0.4\,\mathrm{GeV\,cm^{-3}}$. We remark that these local recoil inputs are held fixed when varying the velocity scale of the gamma-ray emitting region.

Finally, the accepted recoil spectrum is
\begin{equation} \label{eq:dd_accepted_spectrum}
    \frac{dN}{dE_R}(E_R) =\mathcal E\,\varepsilon_{\mathrm{LZ}}(E_R) \frac{dR}{dE_R},
\end{equation}
where $\varepsilon_{\mathrm{LZ}}$ is the tabulated LZ selection efficiency and the exposure is $\mathcal E=2.84$~tonne~yr~\cite{LZ:2026axp}. At fixed masses and splitting, the rate scales as $\rho_\chi \alpha_D \epsilon^2$.

\paragraph{The transition dipole contribution}
The electromagnetic transition responsible for $\chi_2\to\chi_1\gamma$ also mediates endothermic scattering through photon exchange. To assess this contribution, we consider the low-energy interactions of Eqs.~\eqref{eq:vector} and \eqref{eq:dipole}.
In Eq.~\eqref{eq:width}, the lifetime fixes $|\mu_{12}|$, while its relative sign with respect to $g_D\epsilon$ determines the sign of vector-dipole interference.

\begin{figure*}[ht!]
    \centering
    \includegraphics[width=\linewidth]{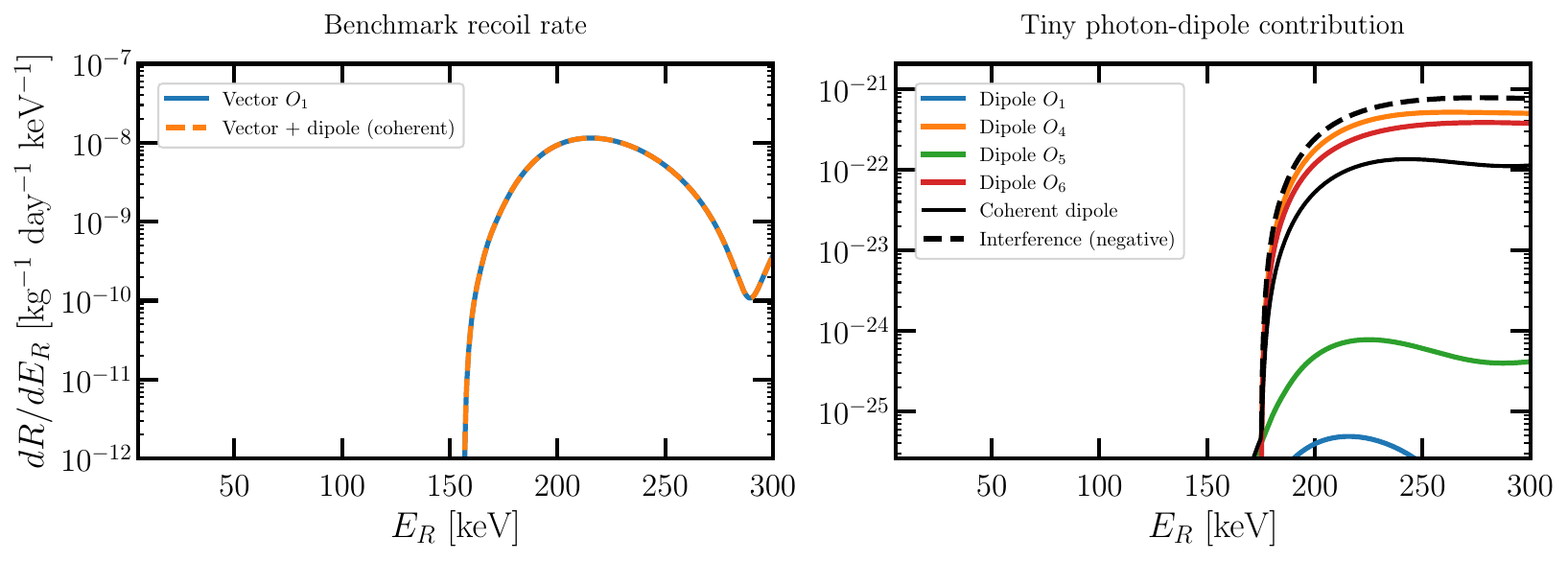}
   \caption{The recoil rate as a function of recoil energy $E_R$ for our main benchmark parameters. The left-hand side panel shows a comparison between the pure dark photon vector contribution ($\mathcal{O}_1$) and the total rate including coherent dipole contributions, showing no appreciable difference. The right-hand panel illustrates the individual electromagnetic dipole operator contributions ($\mathcal{O}_1, \mathcal{O}_4, \mathcal{O}_5, \mathcal{O}_6$), their coherent sum, and the negative interference term, highlighting their extreme suppression by over 10 orders of magnitude relative to the vector component.}
    \label{fig:dd_comparison}
\end{figure*}

At the nucleon level, the photon couples through
\begin{align}
    \langle N'|eJ_{\rm EM}^\mu|N\rangle = & \nonumber  \\ 
    e \, \overline u_N' \bigg[ F_1^N(k^2)\gamma^\mu 
    &+\frac{iF_2^N(k^2)}{2m_N}\sigma^{\mu\nu}k_\nu \bigg]u_N, \label{eq:dipole_dd_nucleon_current} 
\end{align}
where $N=p,n$ and $k=p_N'-p_N$. Contracting this current with the transition-dipole current supplies the photon propagator $1/k^2\simeq-1/q^2$, with $q=|\boldsymbol k|$. The matching used below retains the static nucleon form factors,
\begin{equation}
\begin{gathered}
    Q_N = F_1^N(0), \quad g_N = 2[F_1^N(0) + F_2^N(0)],\\
    (Q_p, Q_n) = (1,0),\quad (g_p, g_n) \simeq(5.59,-3.83).
\end{gathered}
\end{equation}

At leading order in the matching, the reduced non-relativistic amplitude can be written as
\begin{equation}\label{eq:dipole_dd_nr_amplitude}
    \mathcal{A}_N^{\rm dip} =\sum_{i=1,4,5,6}c_i^{N, {\rm dip}}(q)\mathcal{O}_i,
\end{equation}
where the operator basis is given by
\begin{align}
    \mathcal{O}_1 & = \mathbf{1}_\chi \mathbf{1}_N\,, \\ 
    \mathcal{O}_4 &= \mathbf{S}_\chi \cdot \mathbf{S}_N\,,\\
    \mathcal{O}_5 &= i \mathbf{S}_\chi \cdot \left( \frac{\mathbf{q}}{m_N} \cross \mathbf{v}_\perp \right),\\
    \mathcal{O}_6 &= \left( \mathbf{S}_\chi \cdot \frac{\mathbf{q}}{m_N}\right) \left( \mathbf{S}_N \cdot \frac{\mathbf{q}}{m_N}\right)\,.
\end{align}
We orient the momentum transfer $\mathbf{q}$ so that $\mathbf{v} \cdot \mathbf{q} = - q^2 / (2 \mu_{\chi N}) - \delta$, and define
\begin{align}
    \mathbf{v}_{\rm inel}^{\perp} = & \; \mathbf{v} +\frac{\mathbf{q}}{2\mu_{\chi N}} +\frac{\delta}{q^2}\mathbf{q}, \\
    \mathbf{v}_{\rm inel}^{\perp} \cdot \mathbf{q} = & \; 0 \,,
\end{align}
where $\mu_{\chi N} = m_1 m_N/(m_1 + m_N)$.
In the sign and normalization convention of Ref.~\cite{Eby:2023wem}, the nucleon coefficients are
\begin{equation}
\begin{split}
    c_1^{N, {\rm dip}} & = \frac{e \mu_{12} Q_N}{2m_1}, \\
    c_4^{N, {\rm dip}} & = \frac{e \mu_{12} g_N}{m_N}, \\
    c_5^{N, {\rm dip}} & = -\frac{2 e \mu_{12} Q_N m_N}{q^2},\\
    c_6^{N, {\rm dip}} & = -\frac{e \mu_{12} g_N m_N^2}{m_Nq^2}.
\end{split}
\label{eq:dipole_dd_coefficients}
\end{equation}
In the static approximation for the form factors, the charge coefficients $c_{1,5}^{n, {\rm dip}}$ vanish, whereas the neutron magnetic moment generates nonzero $c_{4,6}^{n, {\rm dip}}$. Retaining the leading vector-exchange operator, the combined coefficients are
\begin{equation}
\begin{split}
    c_1^{N, {\rm tot}}(q)&= \frac{g_D \epsilon e Q_N}{m_{A'}^2+q^2} + c_1^{N, {\rm dip}}(q),\\[2ex]
    c_i^{N, {\rm tot}}(q)&=c_i^{N, {\rm dip}}(q),
\end{split}
\label{eq:dipole_dd_total_coefficients}
\end{equation}
with $i=4,5,6$. Thus, photon exchange contributes a correlated set of operators in addition to the dominant proton charge interaction.

The nuclear responses relevant to the dipole are the charge response $M$, the transverse spin response $\Sigma'$, the orbital-current response $\Delta$, and the mixed response $\Delta\Sigma'$. 
Since the magnetic spin interaction couples only to spin components perpendicular to the momentum transfer $\mathbf{q}$, the operator $\mathcal{O}_4$ initially contains both longitudinal and transverse contributions, but $\mathcal{O}_6$ removes the longitudinal contribution through destructive interference. Ultimately, this gives rise to the essential correlation,
\begin{equation}\label{eq:dipole_dd_longitudinal_cancellation}
    \frac{q^2}{m_N^2}c_6^{N,\rm dip} = - c_4^{N,\rm dip} .
\end{equation}
The longitudinal spin response $\Sigma''$ therefore cancels, while the transverse spin response remains. This is an amplitude-level $\mathcal{O}_4$--$\mathcal{O}_6$ cancellation, rather than permission to discard $\mathcal{O}_6$ before squaring the amplitude. 
The mixed $\Delta\Sigma'$ response also contains interference between the orbital and magnetic terms, proportional to products of $c_5$ and $c_4$. 

Consequently, summing rates calculated with isolated operators would miss both effects. Instead, we calculate the total non-relativistic amplitude for a xenon isotope $T$ as
\begin{equation}
\begin{split}
    \overline{|\mathcal{A}_T|^2} ={}&\overline{|\mathcal{A}_T^{A'}|^2} +\overline{|\mathcal{A}_T^{\rm dip}|^2} + 2\operatorname{Re}\overline{\mathcal{A}_T^{A'}\mathcal{A}_T^{\rm dip*}}.
\end{split}
\label{eq:dipole_dd_vector_interference}
\end{equation}
For unpolarized scattering and the retained leading vector operator (dark photon), the vector amplitude interferes only with the dipole $\mathcal{O}_1$ term, whereas interference with $\mathcal \mathcal{O}_{4,5,6}$ vanishes under the DM spin average. 

The recoil rate per unit detector mass is computed using Eq.~\eqref{eq:dd_rate}, but now taking into account that the squared amplitude has the form $A_{0, T}(q) + A_{1, T}(q)\,v_{\perp, T}^2$, with $v_{\perp, T}^2 = v^2 - v_{\min, T}^2$. Therefore, in addition to the inverse-speed integral in Eq.~\eqref{eq:dd_eta}, the $\mathcal{O}_5$ contribution requires the moment
\begin{equation}\label{eq:dd_eta_perp}
    \eta_\perp(v_{\min}) = \int_{v \geq v_{\min}}d^3v\, \frac{f_{\rm lab}(\mathbf{v})}{v}(v^2 - v_{\min}^2).
\end{equation}
The rate combines $A_{0, T} \,\eta + A_{1, T}\,\eta_\perp$ and is folded with the same detector efficiency and energy response used for dark photon exchange. The number of events is also calculated as discussed above for the total differential cross-section. At fixed masses and $g_D$, the full spectrum contains terms proportional to $\epsilon^2$, $\epsilon\mu_{12}$, and $\mu_{12}^2$. 

For our parameter choices, these corrections are negligible. The radiative lifetime implies
\begin{equation}
\begin{split}
    |\mu_{12}|&\simeq\sqrt{\frac{\pi\hbar}{\tau_\gamma\delta^3}}\\
    &\simeq 2.168\times10^{-11}\,\mathrm{GeV}^{-1} \left(\frac{10^8\,\mathrm{s}}{\tau_\gamma}\right)^{1/2} \left(\frac{353\,\mathrm{keV}}{\delta}\right)^{3/2}
\end{split}
\label{eq:dipole_dd_moment_estimate}
\end{equation}
in a xenon recoil of $248\,\mathrm{keV}$.

Fig.~\ref{fig:dd_comparison} shows the impact of the electromagnetic dipole moment operators on the recoil rate. The left-hand panel compares the pure dark photon contribution (vector $\mathcal{O}_1$) with the total rate including the coherent dipole contributions, showing that the two curves are virtually indistinguishable. The right-hand panel highlights individual dipole operator contributions and their coherent sum, which remain suppressed by roughly $13$ to $15$ orders of magnitude relative to the vector contribution. Consequently, for our benchmark, the dark photon (vector) contribution entirely dominates the recoil rate.

\paragraph{Exothermic recoils and excited-state constraints} 

\begin{figure}[t!]
    \centering
    \includegraphics[width=\linewidth]{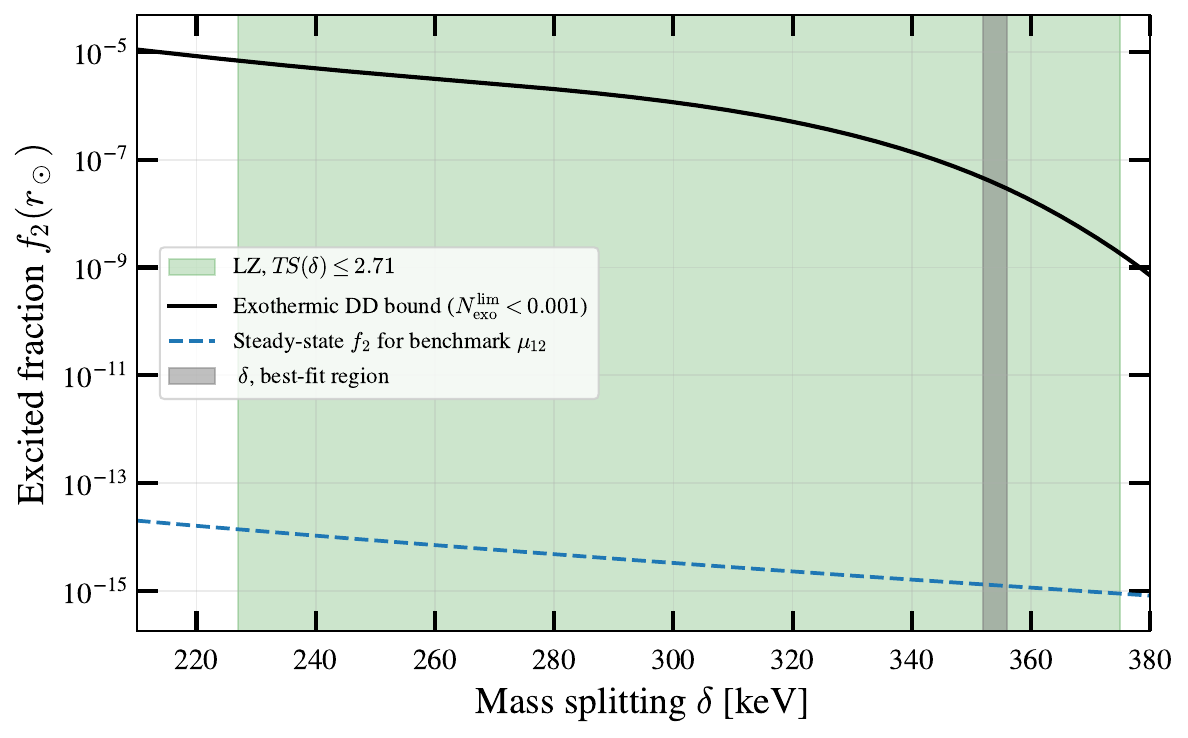}
    \caption{An upper bound on the local excited-state population $f_2$ as a function of the mass splitting is shown in black. The constraint requires the surviving excited-state population to yield a nuclear-recoil signal below the adopted direct-detection limit, $N_{\rm exo}^{\rm lim} < 10^{-3}$. The green region is compatible with the LZ candidate, the gray region sets the best-fit values of $\delta =  352 - 356$~keV, and the dotted-blue curve stands for the steady-state $f_2$ for the benchmark $\mu_{12}$.}
    \label{fig:f2_exothermic_dd_bound}
\end{figure} 

For the reverse transition, $\chi_2 + T \to \chi_1 + T$, the mass splitting $\delta < 0$, so that the minimum speed to produce a nuclear recoil with energy $E_R$ is~\cite{Baer:2026fpy}
\begin{equation}
    v_{\min}^{\rm exo}(E_R)= \frac{1}{\sqrt{2m_TE_R}} \left|\frac{m_T E_R}{\mu_{2T}} + \delta \right|\,.
\end{equation}
Then, the direct exothermic rate at the full density ($f_2 = 1$) is
\begin{equation}\label{eq:exo-rate}
 \frac{d R_{\rm exo}^{(f_2 =1)}}{d E_R} =\frac{\rho_2}{m_2}\sum_T n_T \int d^3v\,f_{2,{\rm lab}}(\mathbf{v})\,v \frac{d\sigma_T^{\rm exo}}{d E_R},
\end{equation}
where changing the sign of the splitting changes kinematics, while $\rho_2/m_2$ sets the incident excited-particle flux. In general, $f_{2,{\rm lab}}$ may differ from the ground-state distribution. After evaluating the negative mass splitting spectrum at the full density, we multiply its accepted count by the \textit{Galactic regenerated} population
\begin{equation}\label{eq:exo-diff-rate}
    \frac{d R_{\rm exo}}{dE_R} \simeq f_2 \frac{d R_{\rm exo}^{(f_2 =1)}}{d E_R},
\end{equation}
and hence, $N_{\rm exo}\simeq f_2\,N_{\rm exo}^{(f_2 = 1)}$. Notably, the expected event number is proportional to the local excited-state fraction for fixed masses, scattering couplings, and velocity distribution.

A particular upper-bound on the surviving excited-state population is set to prevent for exothermic scatterings in the LZ experiment. This is presented in solid black curve in Figure~\ref{fig:f2_exothermic_dd_bound}. This implies $f_2^{\rm lim} \leq N_{\rm exo}^{\rm lim}/N_{\rm exo}^{(f_2 = 1)}$ for our choice of $N_{\rm exo}^{\rm lim} < 10^{-3}$. The green band indicates the region compatible with the LZ candidate using an approximate energy-only likelihood, $TS(\delta) < 2.71$. The gray band denotes the best-fit values for $\delta = 352 - 356$~keV, and the dotted-blue curve stands for the steady-state $\chi_2$ population for the benchmark value of the dipole moment, $\mu_{12}$, confirming that our benchmark choice generate a negligible number of exothermic events. Increasing the lifetime increases the excited state fraction, according to Eq.~\eqref{eq:f2-def}.

\section{Calculation of the excitation cross section}
\label{sec:excited}

For the benchmark we are considering, $\chi_1\chi_1 \to \chi_2\chi_2$ happens in the non-perturbative regime ($\alpha_D m_1 / m_{A'} \gtrsim 1$). To calculate the halo regeneration cross section, we need to solve coupled Schrödinger equations for the initial and final two-body states $|11\rangle$ and $|22\rangle$, obtain the $S$ matrix for each partial wave $\ell$, and find the numerical cross section by summing over the partial wave contributions~\cite{Slatyer:2009vg,Finkbeiner:2014sja, Schutz:2014nka}. 

The dark photon exchange interaction can be described by a central potential with off-diagonal Yukawa terms enabling the $\chi_1-\chi_2$ conversion:
\begin{equation}
    V(r)=\begin{pmatrix} 
        0 & -\alpha_D e^{-m_{A'}r}/r\\
        -\alpha_D e^{-m_{A'}r}/r & 2\delta
 \end{pmatrix}\,,
\label{eq:potential}
\end{equation}
with $r=|\vec r_a - \vec r_b|$ the relative position between the two incoming particles $a$ and $b$. The $2\delta$ term in the potential accounts for the energy cost of producing two excited states.

We thus describe the problem in terms of the relative movement, with center of mass position $\vec R=(m_a \vec r_a + m_b \vec r_b)/M$, where $m_i$ and $\vec r_i$ are the individual masses and positions and $M=m_a+m_b = 2m_1$ is the total mass. The available energy for the relative motion is $E = E_{\rm tot}-\frac{\vec P^2}{2M}$, with the total kinetic energy in the center of mass given by $E_{\rm tot}=2\times \frac{k_1^2}{2m_1}$. The Schrödinger equation for the relative movement reads
\begin{equation}\label{eq:schrodingerPsi}
    \left[ -\frac{1}{m_1} \vec{\nabla}_{r}^2 + V(r)  \right] \vec \Psi (\vec{r}) = E \, \vec \Psi (\vec{r})\,,
\end{equation}
where the relative wavefunction has two components corresponding to the two-body states, $|\vec \Psi(\vec r)\rangle = \psi_1 |11\rangle + \psi_2 |22\rangle$.

Since we are considering a central potential, we can split the radial and angular dependencies of the wavefunctions obeying this equation and solve for each partial wave:
\begin{equation}
    \vec\Psi(\vec r) = \sum_{l=0}^\infty \sum_{m_l=-l}^l \frac{\vec u_l (r)}{r} Y_{lm_l} (\theta, \phi)\,.
\end{equation}
In spherical coordinates, the Laplacian is $\vec \nabla_r^2 = \frac{1}{r^2}\frac{\partial}{\partial r}\left(r^2 \frac{\partial}{\partial r} \right) - \frac{\hat L^2}{\hbar^2r^2}$, with $\hat L^2$ the orbital angular momentum. We can thus solve Eq.~\eqref{eq:schrodingerPsi} for the radial wavefunction at each partial wave:
\begin{equation}
    \left[- \frac{1}{m_1}\frac{d^2}{dr^2} \mathbb{I} + \frac{l(l+1)}{m_1 r^2} \mathbb{I} + V(r) \right] \vec u_l (r) = E \,\vec u_l(r) \,,
\end{equation}
where we have used $\hat L^2 Y_{lm_l} = \hbar^2 l (l+1) Y_{lm_l}$. 

We thus have two second-order equations, with four integration constants to determine. By requiring a finite solution at the origin, $\vec \Psi (\vec 0) \to \vec 0$, we eliminate two constants. We are left with two independent solutions for the radial wavefunctions $u_i(r)$ ($i=1,2$), organized as columns of a radial solution matrix for a given partial wave $U_l(r)$:
\begin{equation}
U_l(r) = 
\begin{pmatrix}
  u_1^{(1)} & u_1^{(2)} \\
  u_2^{(1)} & u_2^{(2)}
\end{pmatrix}\,.
\end{equation}
Defining the dimensionless variables $x=m_{A'}r$, $p_i=k_i/m_{A'}$, and $a=\alpha_Dm_1/m_{A'}$, the equation we solve numerically is therefore
\begin{equation} \label{eq:schrodingerU}
U_l'' = 
\begin{pmatrix}
  l(l+1)/x^2 - p_1^2 & -ae^{-x}/x \\
  -ae^{-x}/x & l(l+1)/x^2 - p_2^2 
\end{pmatrix} U_l \,.
\end{equation}

Any regular solution can be written as $\vec u_l(x) = U_l(x) \vec c$, with $\vec c$ parameterizing the two remaining integration constants. We find these constants from the asymptotic radial solutions. Far from the scattering region, the potential is $V(r\to \infty) \to \rm diag(0,2\delta)$ and the two equations become simply
\begin{equation}
    u_i''+\left[p_i^2-\frac{l(l+1)}{x^2}\right] u_i=0\,.
\end{equation}
The solutions are two decoupled free partial waves which can be written as $u_i(x) = a_i F_i(x) - b_i G_i(x)$,  with $a_i$ and $b_i$ constants and the functions $F_i(x)$ and $G_i(x)$ defined as:
\begin{align}
& F_i(x) = \frac{p_i x \, j_l(p_i x)}{\sqrt{p_i}} \\
& G_i(x) = \frac{p_i x \, y_l(p_i x)}{\sqrt{p_i}}\,,
\end{align}
where $j_l(z)=\sqrt{\frac{\pi}{2z}}J_{l+1/2}(z)$ and $y_l(z)=\sqrt{\frac{\pi}{2z}}Y_{l+1/2}(z)$ are the regular and irregular spherical Bessel functions of the first and second kinds, respectively.

For large $p_i \, x$, we find $F_i(x)\sim \sin(p_ix-l\pi/2)/\sqrt{p_i}$ and $G_i(x)\sim -\cos(p_ix-l\pi/2)/\sqrt{p_i}$. In terms of a phase $\phi_i=p_ix-l\pi/2$, we find the asymptotic solutions
\begin{equation}
    u_i(x) \sim \frac{a_i + i b_i}{2i\sqrt{p_i}} e^{i\phi_i} - \frac{a_i - i b_i}{2i\sqrt{p_i}} e^{-i\phi_i}\,,
\end{equation}
which can be identified as a sum of outgoing and incoming waves. In terms of matrices, we can write $U=F A-GB$ and $U' = F' A - G' B$, which determines the two remaining constants as $A=G'U-GU'$ and $B=F'U-F U'$. We then write the outgoing and incoming amplitude vectors $\vec \alpha_{\rm out} = (A+iB)\vec c$ and $\vec \alpha_{\rm in} = (A -iB)\vec c$. For an incoming state $|11\rangle = (1,0)^T$, we have $\vec c=(A-iB)^{-1}\rm (1,0)^T$. 

The scattering matrix $S_l$ is defined as $\vec \alpha_{\rm out} = S^{(l)} \vec \alpha_{\rm in}$ and can be calculated numerically through 
\begin{equation}
    S^{(l)} = (A+iB)(A-iB)^{-1}\,.
\end{equation}

The asymptotic solutions for the excitation process $\chi_1\chi_1\to\chi_2\chi_2$ are thus given by
\begin{align}
& u_1(x) \sim \frac{1}{2i\sqrt{p_1}}\left(S_{11}^{(l)} e^{i\phi_1} - e^{-i\phi_1}\right) \\
& u_2(x) \sim \frac{1}{2i\sqrt{p_2}} S_{21}^{(l)} e^{i\phi_2} \,.
\end{align}

To calculate the excitation cross section, we need the incident and scattered probability currents which are in general given by $\vec j\equiv \rm Im(\Psi^*\vec \nabla \Psi)/\mu$, with $\mu$ the reduced mass of the state $\mu = m_1/2 \approx m_2/2$. From the asymptotic solutions above, we have an incident plane wave $\Psi_{\rm inc}(\vec r) = e^{i\vec k_1 \cdot \vec r}$, leading to an incoming probability current $j_{\rm inc}=\vec k_1/\mu$, as well as a reflected fraction. The scattered state has a wave $\Psi_{\rm scatt}(\vec r) \sim f_{21} e^{i\vec k_2 \vec r}/r$ that leaves the scattering region, with scattered flux given by $j_{\rm scatt}=\vec k_2|f_{21}(\theta)|^2/(\mu r^2)$. 

Expanding the exponential factors in terms of Legendre polynomials, the solutions of the excitation process are the following normalized wavefunctions:
\begin{equation}
    \Psi_j(\vec r) = \sum_{l=0}^\infty (2l+1) i^l P_l(\cos\theta) \frac{S_{j1}^{(l)}e^{i\phi_j} - \delta_{j1}e^{-i\phi_j}}{2ir\sqrt{k_1 k_j}}\,,
\end{equation}
such that
\begin{equation}
    f_{21}(\theta)=\frac{1}{2i\sqrt{k_1 k_2}} \sum_{l=0}^\infty (2l+1) P_l(\cos\theta) S_{21}^{(l)}
\end{equation}

The excitation cross section is defined as the scattered flux through a solid angle $d\Omega$ over the incident current per unit of area, so we have
\begin{equation}
    \frac{d\sigma}{d\Omega} = \frac{r^2 \, \vec j_{\rm scatt} \cdot \hat r }{\vec j_{\rm inc}} = \frac{k_2}{k_1}|f_{21}(\theta)|^2\,.
\end{equation}

Therefore, from the solution of the Schrödinger equation~\eqref{eq:schrodingerU}, we can calculate the excitation cross section as a sum of partial wave contributions:
\begin{equation} 
\sigma_{11\to22}(v_{\rm rel})
=\frac{\pi}{k_1^2}\sum_{\ell=0}^{\infty} 2w_\ell(2\ell+1)\bigl|S^{(\ell)}_{21}\bigr|^2\,.
\label{eq:sigma11to22}
\end{equation}
The factor $2w_l$ accounts for undistinguishable particles in the initial state, with $w_l = 1/4$ for even $l$ and $w_l=3/4$ for odd $l$, coming from the fact that the wavefunction must be anti-symmetric for identical particles~\cite{Baek:2021yos}. For the orbital part, we have $Y_{lm_l}(-\vec r)=(-1)^l Y_{lm_l}(\vec r)$, which is (anti-)symmetric for (odd) even $l$, so the allowed spin state for even $l$ is an anti-symmetric singlet state and the allowed spin states for odd $l$ are three symmetric triplet states. 

\end{document}